# Fusion Driven Transmutation of Transuranics in a Molten Salt

*Joshua Tanner[1], Ales Necas[2], Sydney Gales[3], Gerard Mourou[4], and Toshiki Tajima[1,2]*

[1]*University of California Irvine, Irvine CA, USA.*

*E-mail address: jetanner@uci.edu*

[2]*TAE Technologies, Foothill Ranch CA, USA.*

[3]*Institut de Physique Nucleaire d' Orsay, IN2P3/CNRS and Université Paris-Saclay, France.*

[4] *Ecole Polytechnique, Route de Saclay, Palaiseau, France.*

A first set of computational studies of transmutation of spent nuclear fuel using compact tunable 14 MeV D-T fusion driven neutron sources is presented. Where we study the controllability, time evolution, as well as effects of spatial distribution of the neutronics in the transmutation in the subcritical operations regime of a transmutator, in which our neutron sources are small, distributed, and can be monitored. Source neutrons are generated via beam-target fusion whereas a deuteron beam is created by laser irradiation of nanometric foils, through the Coherent Acceleration of Ions by Laser (CAIL) process, onto a tritium soaked target. This can be accomplished using relatively cheap fiber lasers terminating onto small scale targets which makes possible the use of multiple tunable and distributable neutron sources. This source is then combined with a molten salt core whose liquid state allows: homogeneity by mixing, safety, in-situ processing, and monitoring. Such a source and molten salt combination allows for the introduction of rapid feedback or feedforward control of the system's operation that have not previously been considered. This encourages an investigation with the aid of AI into new spatial and operation control strategies as done here.

Keywords: Fusion, Artificial Intelligence, Neutronics, Radioactive Waste Disposal, Molten Salt

# 1 Introduction

The utilization of neutrons of fusion has substantial roots in the fusion research. In a fusion reactor, in which the energy production is the ultimate goal, a substantial energy fraction of neutrons may be directed to its' energy conversion, where the Q-value (the energy production divided by the input energy to trigger fusion) be greater than unity. Meanwhile, if the goal is not to make a fusion reactor, this constraint may be relaxed. Here we consider a transmutator of radioactive spent nuclear fuel (SNF) driven by fusion-derived neutrons. In such a transmutation, therefore, the Q-value condition (Q > 1) is not the primary constraint. Many authors suggested [1–6] using deuterium–tritium (D-T) fusion neutrons to transmute SNF. Besides the easier threshold not to reach Q > 1 condition, the fusion neutron energy that exceeds MeV plays an important role in transmuting minor actinides MA (as part of SNF), where MA's are a subset of the transuranic elements (TRU) are the longest lived and most toxic elements in SNF. There have been plenty of beam-driven fusion reaction schemes. The role of the beam may be considered as a more straightforward access to its' directed energy to induce high energy thresholds of nuclear reactions both of triggering fusion and transmutation of nuclei, circumventing the large fraction of thermal plasma population. These include: the beam-fusion [7], the beam-driven muon fusion [8]. In addition, there have been a class of works on the fusion-reactor-generated neutrons [9,10] that could pave the way for transmutation. There is another class of neutrons that are created by high energy proton/deuteron beams in a process called neutron stripping and spallation [11–13].

As mentioned above, the fusion neutron energy (such as 14 MeV neutrons from the D-T fusion reaction) has an advantage in transmutation. The reason is the following. As shown in Fig. *1*, the cross section of neutron capture by Americium (Am-241) is exceeded by fission cross section if the energy of incident neutrons is sufficiently high (such as MeV), as with D-T fusion-produced neutrons (and its immediate slowdown). In addition, this helps to overcome small dips in the fissionability that would otherwise be steadily increasing from an increasing nucleon count as seen in Fig. *2*. The result is the transmutation of these nuclides, such as Americium and other Minor Actinides (MA), that accumulate in these fissionability dips from repeated captures of the uranium fuels. This would diminish the loss of potentially fission causing neutrons needed to continue the secondary fission chain reaction. A properly rebalanced system would not only stop the accumulation but remove existing MA which are the heart of the nuclear waste issue. Additionally, any nuclide that does undergo neutron capture instead of fission would also move more precariously up the atomic ladder and be made more readily fissionable by the same principals. This encourages the finding of a driving fuel that is the result of these captures above the original waste in nucleon count to prevent the further production of waste by continually inserting nuclides from below. This insertion from below is often done with power focused fission process such as traditional nuclear reactors and other planned fuel sources such as thorium reactors.

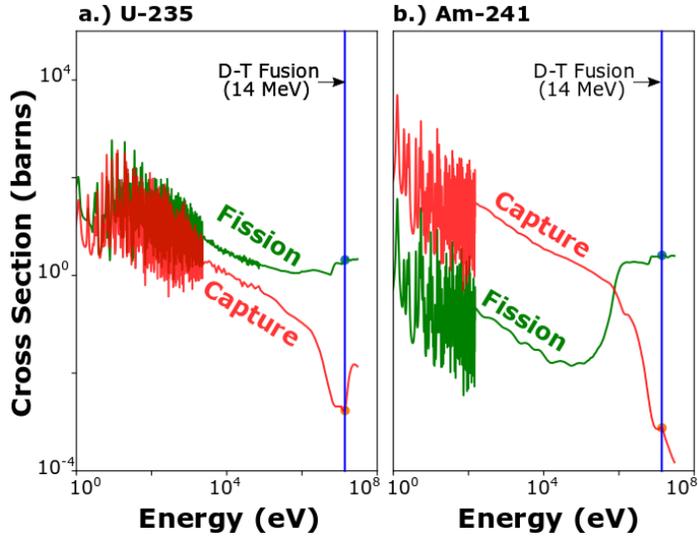

Fig. 1 Comparison of cross sections of the fission induced by a neutron and the neutron capture by $^{235}$U and $^{241}$Am

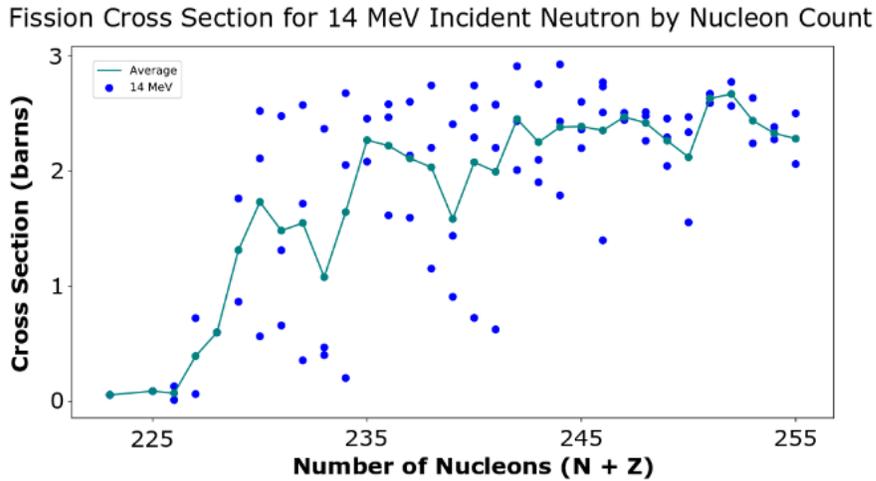

Fig. 2 Increasing fission cross section showing ease of fission following nucleon count. Uses nuclides with greater than 224 nucleons from ENDF/B-VIII.0 library with 900K temperature profiles.

In the present paper we present three main pillars of the principles of how best we operate the D-T fusion-driven neutrons to induce transmutation of transuraniums as presented in [14]. The first pillar of the idea is the distributed use of an innovative laser-driven neutron sources [15,16]. This distributed nature of the source contributes to the safety and flexibility of the operation as well as the cost. The second major feature of our system is to induce transmutation in a molten salt [17] with the TRU's dissolved directly into the salt. This liquid operation, as opposed to the usual solid rod operation of nuclear reactors, has the following distinct advantages: liquid operation allows much wider stability (and thus again safety) and flexibility, along with the convergence of the operational timescales, which also helps in our system. The third major innovation in our operation is the ability of feedback control and its operation by using the artificial intelligence (AI) such as genetic algorithm (GA) [18,19] and neural network (NN) [20–24]. Because our system operates in the molten salt liquid, we converge many time scales into somewhat overlapping time scales of various physical, chemical, and operational processes. In addition, the introduced liquidness allows many of the feedback processes are permitted and transparent (such as the laser or other photon diagnoses are accessible). Note also that many physical processes are too fast for human interventional time scales and only accessible via electronic systems of detection, processing, and feedback controlled computational and transmission time scales. In other words, our fundamental three elements of the major features of our transmutator are intertwined.

We discuss our motivation and the basic ideas in Sec. 2, while in Sec. 3 we survey the methods used in the generalized simulation of this transmutator. Our spatial simulation works by source controlled spatially shaped thermal insertions are shown in Sec. 4. We further discuss temporal aspects through active processing control of a molten salt transmutation by artificial intelligence in Sec. 5. The conclusions are collected in Sec. 6.

## 2  Motivation

The climate crises is the largest and most urgent problem we face today. The climate crisis is also fundamentally a problem of priorities. This problem stems from humanities' over focus on the means of production at the neglect of how to process the waste created from that production. The brunt of the climate issue also revolves around energy production, typically through their emission of greenhouse gases. However, while within nuclear energy we already have and use a well-developed source of power that seems to avoid the problems of an enormous carbon footprint. It unfortunately has other issues that of: Weapons proliferation, Safety, and Nuclear Waste. These issues typically make it an often debated subject on whether it really is the solution that we need, even while it is already in significant use.

Nuclear power currently generates more than 10% of the world's electricity through about 440 reactors operating in nearly 30 countries [25]. This results in an ongoing and significant amount of waste being continuously generated in an area of high dependence and already significant structural investment. Nuclear power generation currently produces a stream of radioactive Spent Nuclear Fuel (SNF), with about 10,000 tons created each year according to International Atomic Energy Agency (IAEA) reports [25].

While how the world approaches power generation in the future is not the focus of this work, it is clear that if nuclear power does play a part the generation of waste would need to be addressed. Yet even if nuclear power is not a significant part of our power generating means in the future, there is already a large quantity of waste that should not be ignored. Worldwide SNF inventory is approximately 300,000 metric tons which has developed from almost 70 years of past power generation [25]. Waste management sufficient to handle at least this preexisting waste burden in concert, or even outside the scope of energy production, should then become a necessity.

To date, other than a deep earth burial of the SNF, there are no well-established long-term approaches that are available to dispense these radioactive materials. Deep geological storage, while considered safe, has several issues that has led to many started and canceled projects but only 1 existing deep geological storage that still does not yet store High Level Waster (HLW) at Onkalo (in Finland) after decades of attempts. Most importantly deep geological storage is not a means of reduction only a means of storage away from likeliest danger. Transmutation is the only means of reducing waste. Additionally, even in the case of only partial transmutation it can be combined deep geological storage with smaller hurdles for greater effect. The partition and transmutation (P&T) waste management method has the goal to eliminate 99.9% of the TRUs. These goals if technically feasible and economically viable present a realistic and safer alternative to the geological repositories in order to eliminate HLW.

Even though the class of highly toxic long-lived radioactive components of SNF (transuranic elements) is a minority, their toxicity is particularly high. However, by following the P&T management we greatly reduce the radiotoxicity of SNF. This motivation for transmutation may be seen in Fig. *3*. Overall, the P&T of the SNF allows 100 times volumetric reduction and 1000 times duration reduction of the storage facility of the remaining waste (mainly fission products) [26].

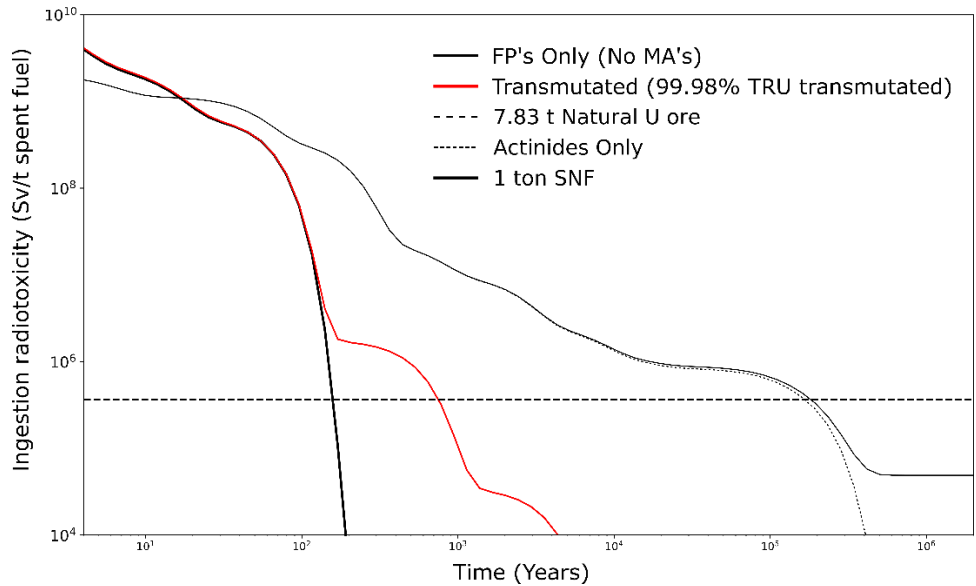

Fig. 3 The ingestion radiotoxicity for 1 ton of the following: Spent Nuclear Fuel, Fission Products of SNF, Actinides of SNF, and scenario of 99.98% of all TRU's Transmuted. Reference line given for 7.83 tons of natural Uranium ore which would be needed to process into 1 ton of Uranium fuel. As shown the radiotoxicity of the FPs dominates the total radiotoxicity during the first 100 years. Long-term radiotoxicity is dominated solely by actinides, mainly plutonium and americium isotopes.

The goal of this research is to suggest a method to transmute radioactive transuranic elements and thereby to reduce the overall radiotoxicity of the SNF. This is the backend of the nuclear fuel cycle, in contrast to the frontend (mining, enrichment, fuel fabrication) which is the "energy production" part of the nuclear fuel cycle. As human activities globally grow, so does the importance on the backend science to dispense the SNF waste associated to this increased energy consumption and accumulated waste. Here we sketch some of the preliminary modest results for a proposed Transmutator using fusion driven neutrons as described here [14]. This includes a novel fusion driven neutron source, a molten salt fuel carrier, and a proposed AI control system working together to perform transmutation of nuclear waste.

## 2.1   Fusion Driven Neutron Source

This fusion driven transmutation is being explored following recent developments of a novel neutron source [14] that utilizes laser driven acceleration techniques to drive fusion neutron production in a compact, relatively cheap, and production rate tunable way (directly tunable in both intensity and timing by control of the driving laser). The source neutrons are generated via beam-target fusion whereas a deuteron beam is created by laser irradiation of nanometric foils through the Coherent Acceleration of

Ions by Laser (CAIL) process. This relatively low deuteron energy is projected onto a solid or gaseous catcher target saturated with tritium and the resultant neutrons are catapulted by fusion and eventually by secondary fission processes. This is driven by a large array of femtosecond pulsed fiber lasers with high-repetition and high efficiency coherently added CAN fiber lasers [27]. The small size of the source, combined with its relatively cheap construction end point of a fiber, make the use multiple sources and their distribution the most important difference spatially between the laser driven source and the single source single input site traditional ADS system. As far as this author is aware no other system considers a controlled distributed set of neutron sources, and as a result this source has the largest significance on this study.

## 2.2   Molten Salt

Another central theme of our transmutator is to operate in the liquid state by dissolving TRUs in a molten salt (e.g. FLiBe, FLiNaK). Such operation allows for a real-time passive and active monitoring and feedback and consequently control of the TRU refueling, reprocessing and criticality. This is then coupled with the monitoring of the distributed neutron sources. Monitoring in real-time allows for a flexible and an optimized operation.

The liquid operation enhances safety and efficiency as it allows to employ active and passive laser monitoring. Additional passive safety could be provided by using a frozen plug on the bottom of the transmutator and allowing it to melt in an event of undesired core temperature increase and discharge the transmutator content into a tank containing neutron absorber, e.g. boron, and simultaneously freezing and encasing the content of the transmutator core thus mitigating dispersal of radiotoxic materials. As a consequence of using molten salt, the pressure in the vessel is close to atmospheric thus in an unlikely event of a catastrophic rapture an explosion and dispersal has low probability. Furthermore, due to the molten salts high melting temperature (~300 °C) the content would quickly freeze if no longer actively heated.

A further advantage of the liquid Transmutator is the removal of the fission products (FP) and refueling while the Transmutator is operating without requiring to shut down. The molten salt and TRU refueling can be done jointly as a mixture or separately whereas the TRU fuel is injected as a pellet or a gas. The removal of FP depends on their chemistry. The noble gases (e.g. xenon, krypton) can be removed online by in-situ by helium injection [28], the remainder composition of the molten salt, FP, and actinides can be partitioned to the pyrochemical reprocessing performed off-line whereas FP are removed and actinides and molten salt are returned for further transmutation. It would also be possible to extract energy from the molten salt by pumping a coolant (such as CO2 or non-alkaline-liquid) through the transmutator core in lieu of pumping out the mixture of molten salt, TRU, and FP into an external heat exchanger. This would help to minimize the presence molten salt and fuel in the external circuitry.

We note that there exist efforts in transmutation R&D is based on possible burning of TRU in the Next generation of Fast Breeder Reactor [29], MOSART project [17] or in Accelerator Driven Systems (ADS) which consists of 100s MeV to GeV class proton superconducting linear accelerator (such as MYRRHA [30] with 600 MeV, 2-4 mA) coupled to subcritical core reactor loaded with TRU as fuel elements. Other approaches employ fusion-fission hybrid technology [31]. We have introduced the rationales and a set of new operations and technologies that accompany the present approach.

The molten salt advantages come from several points, which include its liquidity, laser transparency, its low neutron capture cross section, and most importantly its added safety. The aqueous mixture allows for keeping the whole of the system in a near homogeneous mixed state [30] even if the reactions taking place are not uniform throughout the system. Combined with a high vaporization temperature, additional thermal displacements through convection, and the negative temperature coefficients of, FLiBe and FLiNaK, the molten salts being considered here [5–7,32] safety is increased. This is accomplished through easier to control density perturbations and the easier prevention of voiding that must be accounted for with much greater care in the widely used solid fuel systems.

Additionally, monitoring [33] through the optically clear salts provide a possible low latency feedback vital for short time scale control. While modification of the molten salt solution through removal of fission products (FP) can be done in parts [5–7,30,32] and possibly in-line as the liquidity and homogeneity allows for partitioning of batches without major disruptions in the overall process. Finally, there is a decrease in the typical density for fissile targets from solid fuels, this density can also be varied to a degree by adjusting the concentration of dissolved nuclear waste. The lower density, while not significant, changes the typical time scales of action for prompt and high energy neutrons while the tunable laser driven source with possible optical monitoring vastly increase the response time for a control system. This action response cycle is orders of magnitudes faster than for traditional nuclear systems, which are typically done by thermally or thermal neutron speed relevant actions only. The "thermal" processes used include shifting capture to fission cross section ratios by: neutron moderation, mechanical addition of thermally capturing isotopes, Doppler broadening in the resonance regions of the thermal sections of the cross sections, and density controls through cooling. These could be supplemented by much faster response by electrical and computational control of the source, which would now be possible.

### 2.2.1 Homogeneity

Homogeneity achievable by mixing of a molten salt is the most important spatial difference as compared to solid fuel systems. Artificial heterogeneity allowing for advantages as control through selective neutron energy fissioning/capture ratios can be accomplished using multiple coupled tanks with differing fuel concentrations such as a series of annular tanks or distributed modular cells. Multiple coupled tanks could extend controllability, efficiency and even change overall size significantly. Additionally, a bi-stable approach can be used where a breed burn zones are created by a heterogeneous core arranged either mechanically or through depletion development as in a traveling or standing wave reactor. A multiple tank design could also negate the concept of a 'blanket' because of the ability to distribute multiple sources no longer needing a central point would make the blanket simply non source activated tanks on outer edges of activated source regions. Alternatively, a monolithic single tank design could still be desired for simplicity and reliability. The single tank could use of such concepts as using the lower activity a distance from active source being cooler to keep the molten salt as a solid for less chemical and structural stress at the outer wall. These possibilities are certainly open and more easily allowed by the nature of the source and the molten salt so are worth mentioning but are diverging and beyond the scope of a generalized study.

## 2.3 AI Control

Because of the large parameter space that the features of this transmutator allows, but were left open, much of this study revolved around the integration of AI as a tool to optimize and search for solutions. In essence there are two types of AI's involved here. The primary one that is currently used is the Evolutionary Algorithm (EA). This is a class of heuristics used to explore large parameter spaces to find optimized results. The other type is the Neural Net (NN) which takes a trove of previously explored parameter space to "learn" by building up a "weight mapping". The Neural Net can then, with enough learning, quickly (and accurately) "solve" the very hard problems needed without actually "solving" the model equations behind them by mapping a series of inputs onto predicted outcome. Done by following through a much simpler set of functions adjusted by their "weight mapping". NNs are important here because of this speed from not having to solve these complex equations operating in short time scales similarly as employed for fusion in [22,23] for tokamaks and FRC, respectively. However, because they require a system to already have a previously explored parameter space for their learning library, which does not exist for many aspects of this system, luckily an EA can and are often used to generate this data and train neural nets.

AI optimization methods are particularly useful if the distinction of source neutrons and secondary chain reaction fission neutrons can be accounted for well enough to drive effects relating to their different energies. This would pair with differing isotopic evolution pathways that would result to create a guided pathway that can result in efficient and thorough waste burning. These pathways could for example take advantage of breeding making the loss of some neutrons to capture in fertile waste to increase the chance that they will produce a sufficient number of neutrons to cover this loss from increased fissionability in the future.

# 3 Transmutator Simulation Methods

The Transmutator that prompted this work as shown in [14] was envisioned as an optimized result of the exploration of several concepts related to the impact of a novel source combined with the use of molten salts and the other concepts listed above. This work is the exploration on a generalized transmutator with many parameters considered open to allow the best general understanding applicable to many offshoots rather than a specific design. A generalization was also best served by the simplification of some parameters, when possible, to allow the least amount of interference by factors that could easily change by design. This also allows for the application of similar concepts to other designs beyond a particular device. However, these parameters were often still set to best demonstrate as close to a real case of an entire transmutator with multiple concepts working together and not just the theoretical framework. The tools and methods such as the use of Monte Carlo were left generalized to allow for quick adaption to more specific cases as needed or for use in future works.

## 3.1 Monte Carlo

Because a change in incoming flux and can shift the entire chain reaction pathway by promoting new self-sustaining reactions or push otherwise self-sustaining reactions below sustainability small change in incoming flux can have large changes in resultant reactions and thus the change the outgoing flux of any neutronics system. As a result, every boundary in a neutronics problem acts as an additional nonlinear coupled term in a neutronics problem. Consequently, direct solutions are not typical possible except for the simplest of geometric parameters. Even inhomogeneous conditions that would quickly arise from depletion in any system not actively correcting for this such as with solid fuels would constitute as a boundary quickly limiting the usefulness of directly solving a real world neutronics problem.

The best means of solving real world geometries then becomes Monte Carlo. This is because Monte Carlo can ignore vastly improbable yet computationally expensive branching probability ratios by treating them as the small probability they are. Monte Carlo uses randomness to solve for systems that have underlying deterministic principles. Statistical systems with well-behaved underlying function and large number of samples will strongly converge on a solutions of a mapping of its probability density or its peak by averaging these samples [34]. This allows for solving for a number of the simpler classical dynamics problem of single random neutrons to give a greater picture of a much larger number of neutrons in the highly coupled system.

Monte Carlo is especially well suited for neutronics problem as it was privately developed for this purpose in the 1930's by Enrico Fermi, and later secretly expanded for real world use in the Manhattan Project and finally codified in computer science for this still same purpose by Metropolis [34,35] in the 1950's. The programs used to simulate neutronics through Monte Carlo methods have become a fundamental aspect of any neutronics study and were used here in conjunction with other solvers, linkages, and self-developed tools to determine the neutronics information fundamental to this research.

### 3.1.1    MCNP / ORIGEN

Monte Carlo N-Particle Transport code also known as MCNP is the oldest and most well-known of the neutronics solvers and was developed by Los Alamos National Lab as a result of the Manhattan Project. To evaluate neutronics characteristics and "burnup" of the transmutators' subcritical molten salt transmutation process as a function of time the MCNP and SCALE software packages are used through an internally developed code that prepares and parses information from each in a "linkage" code.

Software such as MCNP and SCALE are already thoroughly validated & verified for accuracy for their particular scope. However, the needed scope for all parts of the setup described in is not fully covered by any individual software available. Fortunately, many individual aspects are well covered by existing software, and by using a linkage code to connect these separate codes, a complete and coherent picture can be obtained. Additionally, as the code would be linked and controlled by a central platform that is editable in process, this allows for optimization to be performed and feedback controls simulated by artificial intelligence.

Neutron transport which can solve for neutron flux is the primary focus of the Monte Carlo N-Particle code suite MCNP 6.2 and is at the heart of many neutronic linkage codes. However, as subcritical operations are envisioned, special care is taken to include the source neutrons and keep them separable from fission produced neutrons. This is also necessary because they are not typically tracked for critical systems, that are the primary focus of MCNP, as they become greatly outnumbered the closer to critical a system becomes. As source neuron energies are significantly different than neutron energies from fission, there are important differences in cross sections as described by that play a significant role in the transmutation process. This lack of tracking or inclusion of source neutron is the reason the ORIGEN-S module in the SCALE software package is used instead of CINDER, MCNP's included depletion solver.

### 3.1.2 OpenMC

OpenMC is a relatively new, open-source, and community-developed Monte Carlo neutron and photon transport simulation code originally developed by members of the Computational Reactor Physics Group at the Massachusetts Institute of Technology starting in 2011. It is capable of performing fixed source, k-eigenvalue, and subcritical multiplication calculations on models built using either a constructive solid geometry (CSG) or CAD representation. The code supports both continuous-energy and multigroup transport. Since this code is open source, its use is not subject to licensing, with no restrictions on modifications, developments and addition of new capabilities.

## 3.2 Geometry

Initial tests were done to find the effect of geometry to establish a reasonable baseline to be used throughout for comparison purposes. An overly symmetric design such as a sphere or slab was not desired as it may obscure possible shape effects while some symmetry and simplicity were desired for computational purposes. As a result, the baseline use of a cylindrical tank was chosen. Additionally, this baseline had to be large enough that the molten salt was the primary interaction medium, and the overall size was not a primary factor, while not so large that it was not in a realistic scope for multiple possible designs. As the D-T fusion neutrons had a mean free path much smaller than spallation neutrons and on the order of ten centimeters depending on the fuel concentrations a tank that was 1 meter in diameter and 2 meters in length for the interior core as seen in Fig. *4* was used to possibly accommodate multiple sources.

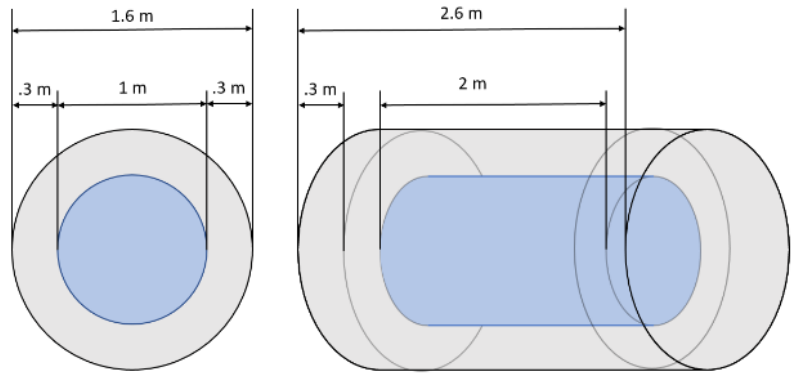

Fig. 4 Schematic of the transmutation tank for the neutronics study.

Many additional spatial parameters could also be considered and optimized by AI but these would fit to more specific conditions rather than general more in the scope of engineering. These optimizable parameters include the use of coupled tanks, exact geometry, and interior design features such as heat exchangers. Thermal Hydraulics could also be considered an important spatial factor, but since anomalous transport would be minimized by the use of a fast spectrum because of the time scale difference between bulk transport and from reaction lifetime from an initial source neutron. As a result, thermal hydraulic were ignored beyond the time scale for mixing for the desired homogeneity as this is a machine specific design focused consideration. An emphasis was instead placed on the relations of the source to transmutator concept particularly the use of multiple spatially distributed sources. This factor is best demonstrated by a spatially shaped reactivity insertion.

### 3.2.1    Outer Wall

A graphite outer wall 30.5 cm thick was chosen as a quick set of simulations showed negligible change in reactions once the wall was more than 2 or 3 times the mean free collision length. Graphite is not required to be the outer wall material and would not be useful for a structural material but was chosen for its high tolerance to radiation (low DPA), low chemical reactivity, and neutron reflectivity that could be used a filler for the bulk shielding. Graphite was also chosen over alternatives such as water because it is an allotrope matching in composition to the diamond skin wall material proposed in [14]. The outer wall material was also confirmed to not represent a significant effect on the overall concept by tests with purely reflective and void boundaries although an observable effect is still present which would require an in-depth study for a completed design.

### 3.2.2 Source Location and Intensity

Input site or distribution is important as any non-infinite medium will have a loss of efficiency due to neutron escape or absorption through a systems outer wall. If a system is subcritical, and thus source driven, this can be minimized by either centralizing the source or making the core large in comparison to the fissioning neutrons mean free path. For this a good approximation of an evenly distributed group of sources is used. The approximation is done through multiple layers of a phyllotaxis arrangements with even spacing in accordance with the total number of sources.

## 3.3 SNF

The origin of waste within SNF is a result of less fissile or reaction dampening materials accumulating in the solid pins used as nuclear fuels until the point that they become no longer effective as a fuel. Because this waste is trapped inside the solid this occurs before a significant amount of the fuel has been used. As shown in **Table *1***, which shows the typical components of 1 ton of SNF from the common operation of a Pressurized Water Reactor (PWR) operated until no longer useable by undergoing a 50 GWd/t burnup before being placed in a cooling storage pool for 6 years. The solid pins with an actinide fuel such as uranium oxide metals ($UO_2$) usually enclosed in a Zirconium Cladding would be used, but about 93% of the original fuel will remain and the rest has become various levels of waste. The two main HLW components are the transuranic (TRU) elements and some Long Lived Fission Products (LLFP such as Tc-99 and I-129). The TRU's in particular, which are produced through consecutive neutron captures and their subsequent decay chains and where half-lives of some isotopes reach millions of years, are responsible for most of the long term radiotoxicity from SNF. For this reason, we will consider the waste to consist of these groups:

- o Unused Fuel: Such as uranium oxide
- o Possible Fuel/Possible Waste: Plutonium

- Minor Actinides (MA) waste: TRUs (except plutonium)
- Fission Products (FP): All remaining fission byproducts.

**Table 1** Composition of 1 ton of SNF from 4.2% enriched UOX fuel in PWR with 50 GWd/tM burnup and 6 year cool down [36]

| \multicolumn{3}{c}{Composition of 1 ton of SNF} | | |
|---|---|---|
| **Element** | **Isotope** | **Mass (kg)** |
| U | 235 | 7.6 |
|   | 236 | 5.46 |
|   | 238 | 922 |
| Np | 237 | 0.7 |
| Pu | 238 | 0.339 |
|   | 239 | 6.09 |
|   | 240 | 2.84 |
|   | 241 | 1.33 |
|   | 242 | 0.85 |
| Am | 241 | 0.502 |
|   | 242m | 0.000902 |
|   | 243 | 0.205 |
| Cm | 242 | 0.000005 |
|   | 243 | 0.000656 |
|   | 244 | 0.0715 |
|   | 245 | 0.00611 |
|   | 246 | 0.00762 |
| FP |   | 51.997 |

The MA portion, and sometimes the Pu portion are used as a baseline for this simulation fuel wise.

## 3.4  Physics of Subcritical Systems

The primary goal of the transmutator is the transmutation or burning of nuclear waste, so determining how much waste is burnt is an important aspect of determining success. How much waste was burned can be seen in the mass of FPs produced, as the loss of mass converted to energy from fission is small compared to the overall mass. With the proposed offline laser based spectroscopy this is possible to determine directly. However typically the mass of many different elements would be difficult to determine actively and a more observational method may also be desired for confirmation. The amount that has been transmuted ($T$) can also be found as a function of the measured amount of thermal energy ($E$) that is released into the molten salt. The transmuted mass is given by the average amount of energy released per fission ($\bar{\epsilon}$) substituted in for the number of fissions that occur ($N$) by combining with the molar atomic mass ($\bar{m}_a$) of what is fissioned

$$T = \frac{\bar{m}_a E}{\bar{\epsilon}}. \tag{3-1}$$

While (3-1) does not include a breakdown of the materials fissioned nor does it include other possible heat contributors, such as beam deposits, radioactive decays and chemical processes, but this does still allows for the quick determination of the expected amounts of TRU's that are burnt in any operating process through traditionally measured means. Which is still useful to see the efficiency through relative measure of waste production through captures to TRUs total burning in terms of scale. This is useable because the average values for the TRU's have a sufficiently narrow range and is not likely to consist of significant amounts of non TRU fission. This TRU specific burn is because the TRU's are also the largest cross sections in this neutron energy range by a significant amount for most operable cases and thus dominate the averages. Additionally, the other contributions to the thermal energy deposited are in general several orders of magnitude smaller than the thermal energy produced by fission. Thus for an expected average thermal energy released by TRU fission between 180 and 220 MeV per fission, and a molar atomic mass average between 238 and 248 u. A generalization of 40.5 ± 4.8 kg per 100 Megawatt year (MWyr) of output thermal power can be estimated as demonstrated below. This can further have narrowed by using expected transmutation targets and excluding ratios that would not be in any likely operating range. SNF would have an expected transmuted mass of 37.4 ± 1.3 kg per 100 MW yr.

To see the relations of input power to output power for determining efficiency. First start working from energy of the output ($E_{out}$) and finding the needed energy of the input ($E_{in}$). Is to take the average energy released by fission ($\bar{E}_f$) divide by the average neutron energy (from fission) ($\bar{E}_n$) and adjust for number of neutrons using normal neutron economy rules (i.e. 1 neutron in = 1/(1-$k_{eff}$) neutrons out except fission is "one step" behind so multiply by another $k_{eff}$). This is done as a substitution for number of fissions so divide by number of neutrons per fission ($\bar{\nu}_f$)

$$E_{in} = \frac{E_{out}\,\bar{E}_n\,(1-k_{eff})\bar{\nu}_f}{\bar{E}_f\,k_{eff}}. \tag{3-2}$$

This can/should be adjusted for actual input in by using adjustment of source to "input" neutrons: Average energy of source neutrons ($\bar{E}_s$), average neutrons per secondary fission ($\bar{\nu}_f$), and average neutrons per source neutron fission ($\bar{\nu}_s$):

$$E_{in} = \frac{E_{out}\,\bar{E}_s\,(1-k_{eff})\bar{\nu}_f\bar{\nu}_s}{\bar{E}_f\,k_{eff}k_{src}} = \frac{E_{out}\,\bar{E}_s\,\left(\frac{1}{k_{eff}}-1\right)\bar{\nu}_f\bar{\nu}_s}{\bar{E}_f\,k_{eff}k_{src}} \tag{3-3}$$

Similarly, with power (E→P) in place of energy and net multiplication (M):

$$P_{in} = \frac{P_{out}\bar{E}_s\bar{\nu}_f\bar{\nu}_s}{k_{eff}k_{src}M\bar{E}_f} \tag{3-4}$$

This allows us to calculate the general effectiveness of the transmutation system through its' input power. Assuming a repetition rate of 100 kHz, the neutron rate per single laser in the source described in [14] is $10^{12}$ n/s. The compactness and scalability of the CAN laser allow us to deploy a large number of units to meet the overall need of the total number of neutrons. Thus, a neutron rate of mid $10^{15}$ n/s can potentially incinerate 10 kg of TRUs a year.

### 3.4.1 Evolutionary Algorithm

There is some issue with calling the algorithm (heuristic) developed here any single specific name. The method names such as: Evolutionary Algorithm (EA), Evolutionary Computing (EC), Genetic Algorithm (GA), Simulated Annealing (SA), and Tabu Search (TS) etc. are not in themselves specific "recipes" but rather frameworks that can overlap. This would be a "hybrid method". An EA could be considered the foundational algorithm used in this hybrid method and is used as the overall description of the AI used for parameter space searching with the GA as its near equivalent as the subpart.

In a broader view Evolutionary Computing (Dortmond 1991) [19] is used as an umbrella term for all algorithms, collectively the "Evolutionary Algorithms", that are inspired by biological evolution specifically meant to cover and merge the independently developed lines of Evolutionary Programming (Fogel 1962), Evolutionary Strategies (Rechenberg 1973), and Genetic Algorithms (Holland 1975) which were all preceded by less defined strategies of Artificial Life (Barricelli 1953), Automatic Programming

(Friedberg 1958), and Artificial Selection (Fraser 1957) and even earlier less referenced work with many additional sub branches (most notably Genetic Programming) later. EA's would be more general of a term than GA, although John Holland made GA a very popular term that is still often used.

In EA's (or ECs or EC Algorithms etc.), an initial set of candidate solutions is generated and iteratively updated. Each new generation is produced by stochastically removing less desired solutions and introducing small random changes. In biological terminology, a population of solutions is subjected to natural selection (or artificial selection) and mutation. As a result, the population will gradually evolve to increase in fitness, in this case the chosen fitness function of the algorithm.

A self developed object based method written in Python was used to allow a mix and match approach of some features of multiple heuristics. Which is often done and considered named by "personal taste" as per [37]. The EA's process as shown in Fig. *5* is:

1. Create a population representing individuals of random guesses of parameter settings.
2. Run individuals through a simulator and score them through a cost function. Where the cost functions setup will determine the optimization goals.
3. After scoring the population select and match the most fit parameters sets by these scores.
4. Prior guesses are then combined by to create a completely new population of parameter settings.
    a. Occasionally some level of mutation is also added to randomly adjust some individuals. This helps prevent results from falling into a local score minima.
5. This new population will be biased towards improvement by taking parts from the most successful of prior guesses.
6. This process is then repeated for a set number of times or until some criteria is met such as no significant improvement is being made.

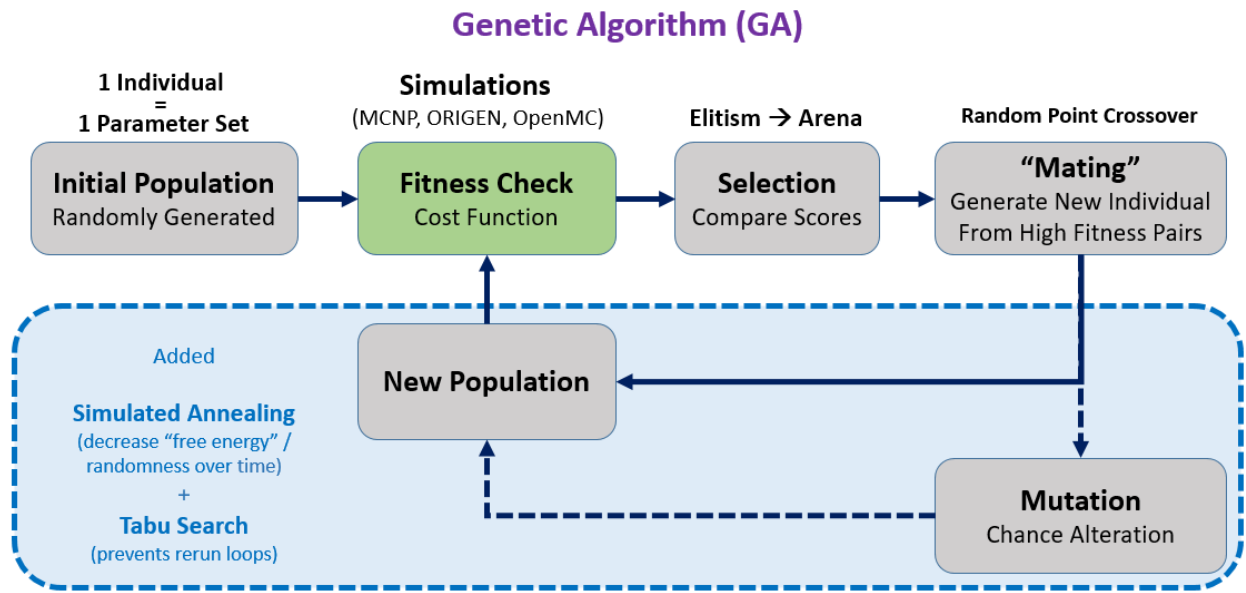

Fig. 5 Process diagram of hybrid EA AI used for optimization.

# 4 Source Controlled Spatially Shaped Thermal Insertion

The biggest danger and limitation for any fission based system comes from thermal deposition. This is because unplanned material vaporization, be it in a structural components, coolant, or fuel, needs to be avoided and become the primary limit to any safe operation. This also extends to secondary effects such as thermal based over pressurization that result in structural breach. Thermal management is especially important to most nuclear systems as overall thermal deposition is usually correlated with the overall power and efficiency.

The operational limit for transmutation is based on the capacity to cool any part of the system. A molten lead (LBE) system typically has a $500 \frac{W_{th}}{cm^3}$ limit which by equation 3-1 would be approximately $190 \frac{kg}{m^3 yr}$ for the volume cooled this way. This is where a major limitation for Aqueous Homogenous Reactors (AHR) utilizing water as its core medium as the water will break down at very low activity levels. It is also not a straightforward matter to utilize this upper limit through the entire tank for a subcritical system. Local power densities become especially important around the sources entry as too high of a local power density can cause state changes resulting in possible issues such as void bubbling which can cause sudden reactivity insertions.

Traditional ADS systems are especially limited by thermal deposition as most accelerators are large and expensive and do not have the option to change input location. While technical improvements can lead to the possibility of higher beam currents, there is an upper limit for single input site efficiency based on thermal management that concentrates on this peak input site. This results in a loss of transmutation efficiency in terms of transmuting the peak amount of material throughout the tank following activity drop off away from this peak input site. Proton spallation alleviates this problem somewhat by greatly increasing the overall initial disbursement area because of the long mean collisional distance from much higher proton energies. Targets made of materials such as tungsten and lead on the scale of a meter long are often used to make spread out distribution of secondary neutrons, that have much smaller mean collisional distances. This disbursement can be further flattened by varying target density or material but is still limited by target design. This contributes to a major efficiency loss from evaporation, and the use of secondary neutrons rather than primary source particles. This can result in losses on the order of a single GeV proton on average producing only 2 or 3 useable ~20 MeV neutrons for an approximately 4-6% efficiency. Compared to D-T fusion produced neutrons this is balanced by obtainable Q factors and importantly here the greater initial spread of neutrons utilizing a higher single source limit in a larger tank.

Sources placed by the wall would indeed suffer more edge-based neutron losses but they allow for a more efficient use of the tank that would otherwise be bounded by the centers highest rate of energy deposition and the limits of cooling at that location. The overall efficiency represented by $k_{eff}$ can also be increased by a higher concentration of fissionable material as fuel to counter this loss somewhat. A placement of multiple sources that balances these efficiencies can yield a higher overall efficiency for each Transmutator.

An effectively random source placement, as shown in Fig. 6a, will often produce a typical centered distribution where there is little benefit from putting sources closer to a wall where efficiency is lowered by neutron escape losses. However, a more even energy deposition allows for more distributed cooling and greater overall power density can be found through iterative AI design as seen in Fig. 6b. Here a Phyllotaxis arrangement of 1000 neutron sources with parameters setting the 10 levels of normalized power is used. This was done as leaving the individual placement of sources as open to the AI tended to increase the parameter space enough to make the required number of simulation runs computationally prohibitive. The AI also had a discretized control on the intensity level of each source for similar reasons.

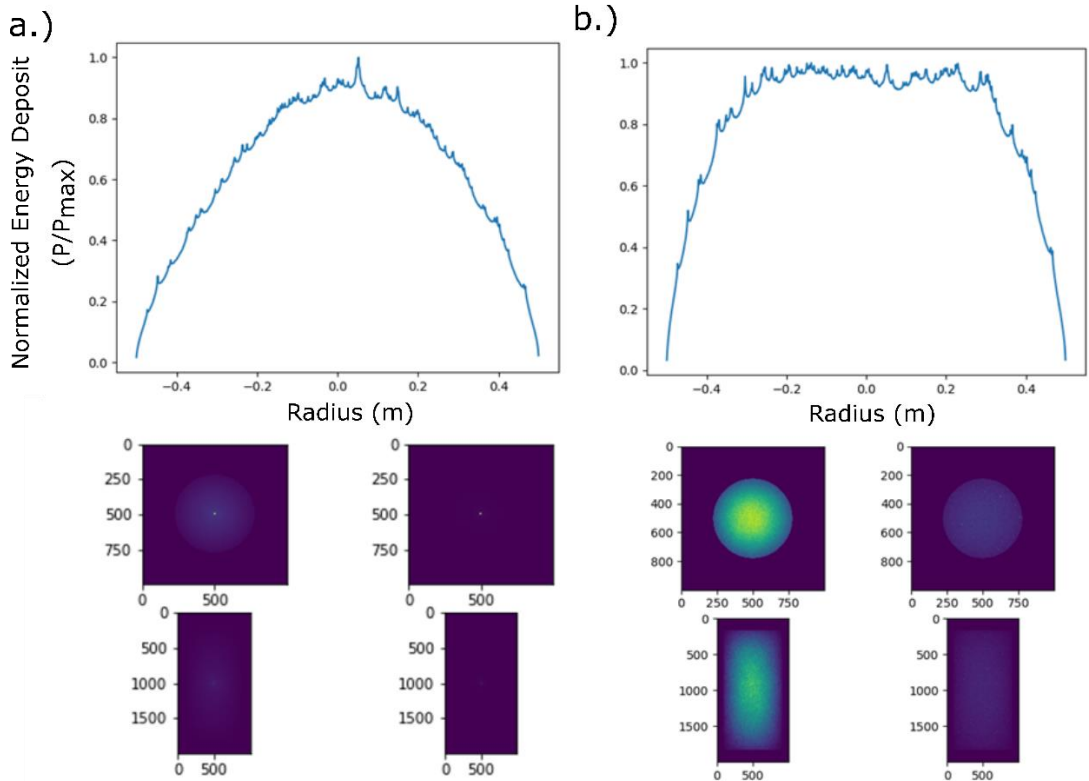

Fig. 6 Thermal deposition profiles for early (a) and late (b) AI optimization results.

These results were found through an EA with a population of 30 at start decreasing by one till a population is achieved for a total of 100 generation steps. Scoring was done by combining the criticality and the difference from maximum for the energy deposition for each part of the tank. In this way losses in efficiency from increased escape is directly measured with the increases in efficiency from a more even power density.

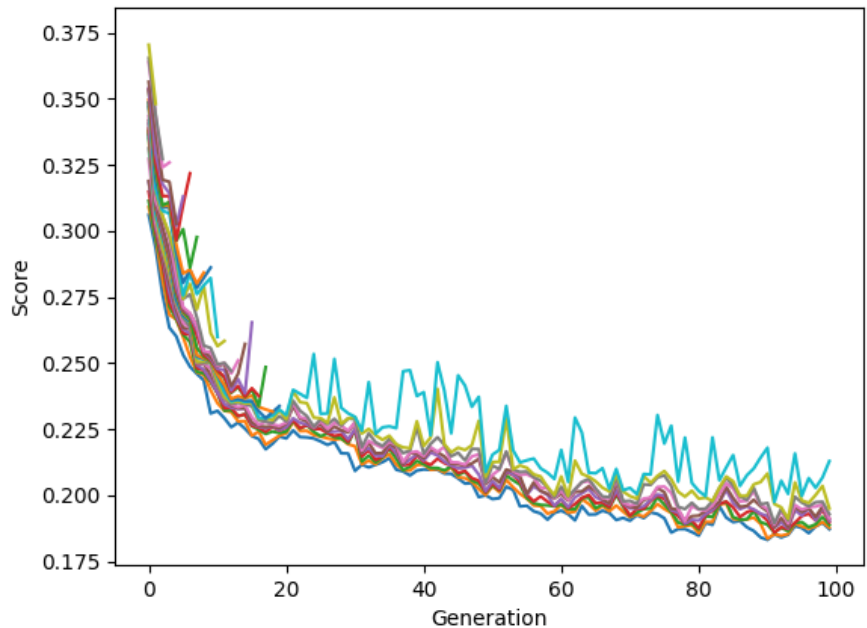

Fig. 7 Scores for all individuals in EA optimization AI population by generation for Fig. 6 a-b.

Additional efficiency can be found by expanding the parameters to include fuel mix and concentration as well as geometric design for separate tanks such as the annular ring style. This would for example allow for a boost to the outer regions of the tank by decreasing the neutron production efficiency (not necessarily the burn efficiency if more desirable but less efficient burn targets such as Am is used) in the most commonly overlapped regions of the center. This thermal insertion shaping can also be extended to safety and operational conditions through active monitoring, allowed by the molten salt, by actively shaping thermal input to bring a current measured system towards an ideal shape to match local thermal dissipation limits.

# 5   Temporal Process Control by AI

The key to making the transmutator sustainable is the adjustment of the concentrations of waste and FP's in the tank. By setting this process to have isotopic concentrations loop over a usable range when more waste is added then something equivalent to a fuel replacement cycle could be created. This can be a steady state operation that allows for the continuous processing of waste. This cycle is not generally straightforward to find as the concentration of isotopes in the transmutator tanks are one of the most significant factors in operation. Molten salt system do allow for many adjustments in-situ, but they are also all separately limited by chemistry and focus on individual or particular groups of elements, the waste also continues to come in the same limited isotopic ratios as initially started.

## 5.1 Operational FP Replacement Process

A major goal for this study is verifying pathways for a no isotope separation waste burner. This is an important factor in the advantage of using a molten salt system in a plausible real-world solution for nuclear waste. Finding such pathways can be accomplished by using AI to show how long-time scale in-situ operational control such as quantity and timing for waste adding and removal allowed by the liquid nature of the molten salt can operate as an efficient waste burner.

For this to be a good candidate as a real-world solution for nuclear waste that can be quickly developed this study was constrained to already existing process such as PUREX to determine what could be added or removed. This is an issue though, since as can be seen in Fig. *8*, using MA as taken directly from typical reprocessing methods as the primary driver for transmutation would be very inefficient. Source neutrons would yield a neutron multiplication effect that would be very close to unity. The $k_{eff} = 0.4$ or $M = \frac{1}{1-k_{eff}} = 1.6$ at a high concentration of 1.6 mol% (~10 wt%) would be well below what is achievable with a $M = 50$ ($k_{eff} = 0.98$) or more of typically considered sub-critical systems. Transmutation of waste directly would then rely strongly on the efficiency of the neutron source to result in significant amounts of transmutations having lost the advantage of the neutron rich environment from neutron multiplication. As an additional note for Fig. *8* there is a nonlinear part where the concentration of MA drops enough to allow the beryllium in the FLiBe to receive a significant portion of the incoming neutrons and undergo (n, 2n) reactions. This would release an increased number of neutrons of sufficient energy to fission causing a boost in criticality. This boost would eventually be lost as the amount of MA to fission drops below what even the additional neutron production can fission.

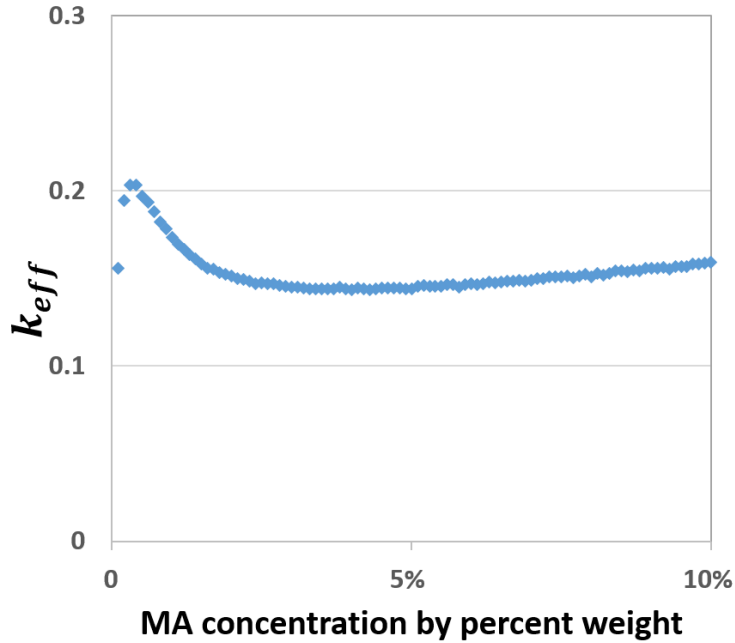

Fig. 8 Criticality of system by fuel to molten salt medium weight with only MA processed from SNF as fuel.

An inefficient MA fuel can be countered by an initial quantity of actinide (normally fissile) fuel, particularly plutonium being initially left mixed with the minor actinides from PUREX. However, our goal is a minor actinide burner and while efficiently using neutrons is good this is still not an efficient MA burner. Minor actinides are very inefficient because even if the first set of nuclides easily fission from fusion neutrons the resultant fission neutrons are not good at maintaining the chain reaction as their energies are more likely to capture than fission again. However, waste is still fertile and if it captures becomes more and more efficient, as shown in *Fig. 9* because of the atomic ladder as mentioned in section 1. This improves the current mixtures efficiency even if after starting only additional MA waste with no Pu is added.

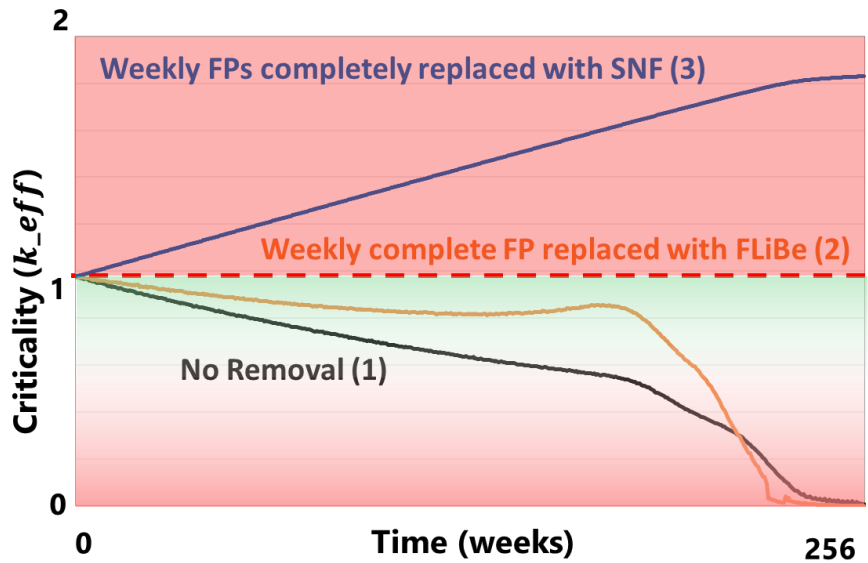

Fig. 9 Comparison of the criticality depending on the operational modes of transmutator: (1) no removal of FP nor input of SNF; (2) weekly removal of all FP replaced with FLiBe; (3) weekly FP removal of all FP replaced by fresh SNF. SNF Loading Fuel is at the 100 MW Thermal Burn Power.

This would normally be hindered by the accumulation of fission products or the reduced overall quantity and thus density of fissile materials, but this can be balanced process wise during operation because of the molten salt. As long as a target trend remains between the trends represented by maximum processable and replaceable with waste or additional molten salt and the trend represented by no processing the fuel can be developed, this would extend to other waste outputs from differing types of plants and inverted trend lines as well. Our goal then becomes allowing for the remaining waste to develop and become a fuel so efficient that the initial Plutonium is no longer needed. This would then turn into a generalized 3-stage process Fig. *10*, the initial being this 'development' phase with a target goal being developing increasingly efficient fuel until the focus can change to efficiently burning waste. After the development phase the transmutator can then be ran in a 'static' phase whereby balancing the in-operation fission product removal and the addition of more minor actinide waste the transmutator runs continuously burning all the waste that is inputted. A 'shutdown' phase could also be considered as the final burn with no additional waste added and TRU quantity reduced until minimal. The 'shutdown' phase while included as markedly different focus will not be covered as the overall importance and effect would be negligible in real world operation as tank contents could be continually combined or operation in the static phase continued for an indefinite time.

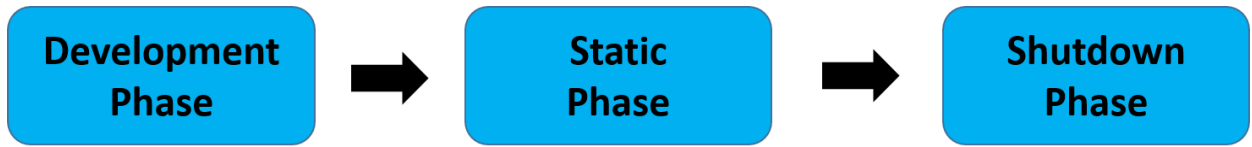

Fig. 10 3 stage transmutation operational plan

This is started with the development phase where, while minor actinide waste burning still occurs, initially most neutrons go to depleting the plutonium. However, enough capture occurs that then develops an increasingly efficient fuel. This primarily derives from Am-241, the normally most problematic waste, capturing a lot of these neutrons resulting in an easier to fission fuel mostly consisting of curium. This is continued until it can be self supporting by the increasing amount of curium. In a solid fuel this would quickly become too inefficient to be worth running or the curium too heterogeneously developed and concentrated and possibly dangerous. However, molten salt can be mixed to continuously homogenize the fuel and can also be processed either continuously during operation or at scheduled stop then start steps. An example of a feedback control scenario, showing (a) criticality change and (b) isotopic evolution with operational scheduling of FP replacement with fresh SNF waste at 50 week intervals (an excessively large period for demonstration purposes) for the development phase can be seen in Fig. *11*.

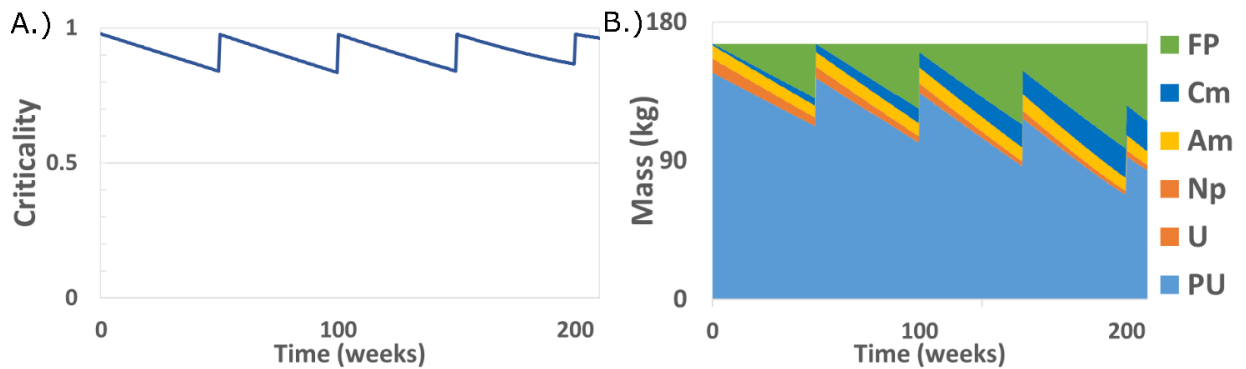

Fig. 11 An example of a feedback control scenario, showing (a) criticality change and (b) isotopic evolution with operational scheduling of FP replacement with fresh SNF waste at 50 week intervals for the development phase.

This operation would also normally be hindered by the steady drop in criticality which is due to fission products capturing neutrons that could otherwise be causing fissions or even being captured and breeding the curium fuel. However, the AI here is determining the ideal amount of processing to perform at these steps to continue operating efficiently. The scoring also keeps in mind other aspects of the system such as power reactivity insertion coefficients for safety purposes. This example is just an extreme on process limiting and step length allowing an exceptionally large power efficiency loss, but in-process removals and more frequent small partition processing can retain better overall power efficiencies as have also done. This is not as important though as this operation type will typically converge on a better or sufficient fuel mixture for reasons such as the higher neutron production per fission of curium than plutonium if other possible neutron losses are controlled, and a fast spectrum is being maintained. Once this mixture is developed it can then be used for the next stage, although it may need to be a combined result from more than one tank to obtain sufficient quantities, as in this case where 9 tanks are combined if all plutonium is then removed.

A developed mixture is then used in the 'static' phase where a similar processing schedule is found by an AI to maintain a consistent balance of creating more of its own fuel and burning all the purely MA waste that is added as shown in Fig. *12*. This allows for burning of an arbitrary amount of waste by continuing the process as long as needed. It should also be noted that the balance is not only just restoring criticality but keeping the integral 'effective' breeding or conversion ratio equal at 1.

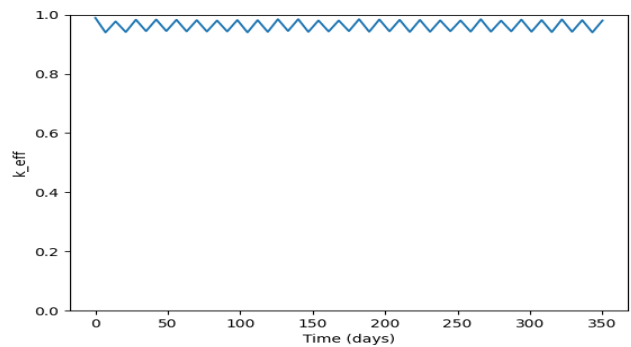

Fig. 12 An example of a feedback control scenario, showing criticality change with operational scheduling of FP replacement with fresh SNF waste at 1 week intervals for the static phase. Transmuted: ~7 kg from operation at $k_{eff}$ = .98 and 1.416 MW$_{th}$ power output.

# 6  Conclusions

In principle the elimination of minor actinide waste through transmutation has been known to be possible, here it is shown that new methods exist that could be used to realize transmutation as a practical solution. These methods in addition to normal nuclear process would be used in a systemic optimization process that improves the effectiveness of the waste elimination. Possible process step can be seen best by the steady state minor actinide burning example where it is shown there are configurations that can, with additional operational control, transmute waste without the creation of more waste.

This example also demonstrates how it may be possible to overcome local minima efficiencies that are sometimes focused on by trying to find power producing processes before considering the waste treatment processes. Typically, this minima comes about as a focus on one of the lower hanging fruits such as the use of more readily available fissionable fuels lower than the waste meant to be treated on the atomic ladder. This process would still allow for an opportunity for the type of further optimizations and exploration of controls that can result in the power multiplier effects envisioned by Rubbia [38] as a future extension. However, power production can now be left as something to be found after treatment of the currently long unsolved problem has already begun and to more completely meet the true objective of waste treatment processes. This would be as a similar technology to the deep waste repository which looks to the management of waste as a true back end solution, with according attention paid to the safety, and efficiency in terms of manageable costs and operation conditions and benefits for its expected time scales.

Through the simulations shown here we can see that by adjusting previous assumptions about limits, by pairing the complementary properties of a liquid system and taking advantage of new cheaper smaller and distributed fusion neutron sources currently being developed, we can take advantage from multiple less explored capabilities. These new capabilities rely on the ability to take advantage of the multi-source placement, source intensity tuning, as well as homogenizing and bulk transport effects in a safer molten salt system that allow for operational scheduling and isotopic evolution in a controlled system. Shown directly is part of a possible initial burn pathways from SNF currently process able from nuclear power plants as in Fig. *11* that are controlled to reach desired isotopic ratios to then allow for steady state burns in the style of Fig. *12*. Steady state can be done through adjustments to the overall operational scheduling that cycle the applied power levels but keep in tack the near linear progression of their isotopic evolutions. This is possible due to SNF's relative insensitivity to the FP and FLiBe carrier ratios at long time scales, except as an overall decrease to their operating power due to moderating effects that are less important to the fissioning of the SNF's initial primary fission driver plutonium. Combined these updated avenues in transmutation could allow for efficient and thorough waste burning tailored to fit any TRU waste in question without creating additional MA waste. Which with further work on the transmutation process in the areas of chemistry, and material science humanity may finally begin to reduce the long standing nuclear waste problem.

# Acknowledgments:

The authors benefited greatly from the discussions, inspirations, teachings, and support from Drs. Yves Brechet, the late Norman Rostoker, F. Carre, Thierry Massard, S. David, Daniel Papp, Bogdan Yamagi, Karoly Osvay, Gabor Szabo, Zsolt Frei, Szabolcs Czifrus, and Maurice LeRoy. The research has been supported by the funds at TAE Technologies and the University of Szeged. In addition, the technical support of the CEA, Institut de Physique Nucleaire Orsay, Ecole Polytechnique, University of Szeged, ELI-ALPs, and the Universite Paris-Saclay. Simulations were conducted chiefly on the TAE Technologies high-performance computing cluster.